\documentstyle[epsfig,color,ulem]{mn}
\title{Toward mode selection in Delta Scuti stars: Regularities in observed and theoretical frequency spectra}

\author[M. Breger, P. Lenz, A. A. Pamyatnykh]
{M.~Breger$^1$, P.~Lenz$^1$, A. A. Pamyatnykh$^{1,2,3}$\\
$^1$ Astronomisches Institut der Universit\"at Wien, T\"urkenschanzstr. 17, A--1180 Wien, Austria\\
$^2$ Copernicus Astronomical Center, Bartycka 18, 00-716 Warsaw, Poland\\
$^3$ Institute of Astronomy, Russian Academy of Sciences, Pyatnitskaya Str. 48,
109017 Moscow, Russia}

\date{Accepted 2006 month day.
      Received 2006 month day;
      in original form 2006 month date}

\begin{document}
\maketitle

\begin{abstract}

Only a fraction of the theoretically predicted nonradial pulsation modes have so far been observed in $\delta$ Scuti stars. Nevertheless, the large number of frequencies detected in recent photometric studies of selected $\delta$ Scuti stars allow us to look for regularities in the frequency spacing of modes. Mode identifications are used to interpret these results.

Statistical analyses of several $\delta$ Scuti stars (FG Vir, 44 Tau, BL Cam and others) show that the photometrically observed frequencies are not distributed at random, but that the excited nonradial modes cluster around the frequencies of the radial modes over many radial orders. 

The observed regularities can be partly explained by modes trapped in the stellar envelope. This mode selection mechanism was proposed by Dziembowski \& Kr\'{o}likowska (1990) and shown to be efficient for $\ell = 1$ modes. New pulsation model calculations confirm the observed regularities.

We present the s-f diagram, which compares the average separation of the radial frequencies ($s$) with the frequency of the lowest-frequency unstable radial mode ($f$).
This provides an estimate for the $\log g$ value of the observed star, if we assume that the centers of the observed frequency clusters correspond to the radial mode frequencies. This assumption is confirmed by examples of well-studied $\delta$~Scuti variables in which radial modes were definitely identified.
\end{abstract}


\section{Introduction}
Recent observational campaigns carried out with earth-based telescopes or space missions concentrating on selected stars have revealed a rich spectrum of radial and nonradial modes covering a wide range in frequencies. This range in frequencies varies from star to star and depends on a variety of factors, not all of which are understood. The comparison of the observations with pulsation models concentrates mainly on the range of frequencies in which pulsational instability occurs as well as mode selection and frequency values for specific modes. In particular, the question of which modes are selected by the star is not solved at the present time. The problem is also a question of amplitude size: we cannot distinguish between stability and oscillations with amplitudes below the level of detectability. Consequently, extensive observational campaigns are needed to lower the observational threshold.

\begin{figure}
\centering
\includegraphics[bb=35 50 750 500,width=85mm,clip]{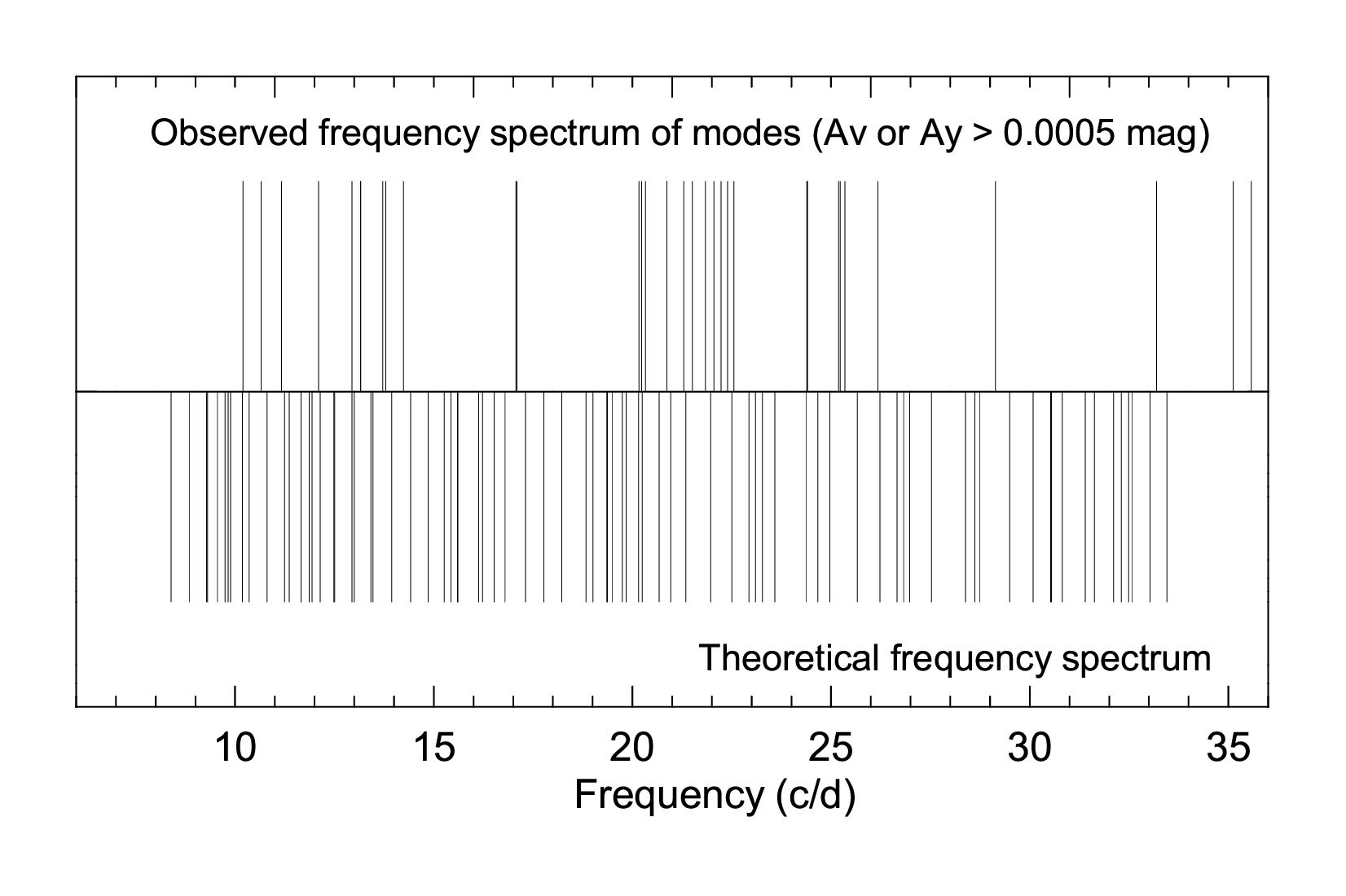}
\caption{Observed and theoretical frequencies of FG Vir. Observed frequencies with amplitudes higher than 0.5 mmag are compared to the theoretical frequency spectrum of unstable modes with $\ell$ = 0 to 2. The rotational splitting of modes for an equatorial velocity of 62.5 km~s$^{-1}$ was taken into account. The predicted spectrum is much denser than observed.}
\end{figure}

The question arises of whether the mixture of the excited radial and nonradial modes shows frequency peaks which are essentially randomly distributed over the range of unstable frequencies or whether they tend to form groups. An example of the latter is the regular spacing found in high-order nonradial pulsation (the asymptotic case), as detected for the Sun and white dwarfs.
The main-sequence and post-main-sequence $\delta$~Scuti stars in the classical instability strip, on the other hand, are nonradial and radial pulsators oscillating with low-order p (and g) modes. The frequencies of pulsation are not known to be regularly spaced. In Fig.~1, the observed frequencies of the  star FG Vir (Breger et al. 2005) are compared with the theoretical frequencies (model used by Breger \& Pamyatnykh 2006). In this figure we have used only the observed modes with photometric amplitudes $\geq$ 0.5 mmag in order to concentrate on $\ell$ = 0 to 2 modes. The comparison shows that a mode and/or amplitude selection mechanism is active: not all the theoretically predicted frequencies are seen. While it is possible to compute whether a mode is unstable or not by means of linear theory (e.g., Stellingwerf 1979 for radial modes), the question of the selection of which overtones are excited to observable amplitudes still remains to be solved by nonlinear computations (e.g., Stellingwerf 1980).
For most of observed modes with small amplitudes, mode identification in terms of the modal quantum numbers
($n$, $\ell$, $m$) is not yet possible. This makes the discovery of the mode selection mechanisms more difficult.

In this paper we examine the frequency distribution of mostly nonradial modes in a number of well-observed $\delta$~Scuti stars in order to search for regularities.

\section{Observational clues}

\subsection{Distribution of detected frequencies in 44 Tau}

Extensive campaigns of 44 Tau covering five observing seasons have led to the detection of 49 frequencies (Breger \& Lenz 2008). The
distribution of these frequencies is shown in Fig. 2. It demonstrates that the detected
frequencies are not distributed at random, but that some semi-regular spacing exists over a very wide range of frequencies. A hint to the
origin of the pattern is given by the examination of the nature of these frequencies: only in the 5.3 to 12.7 cd$^{-1}$ do we find independent
frequencies and all the detected frequencies outside this range have been identified as combination frequencies, a$f_i\pm$b$f_j$, or harmonics.

The frequency region, in which the independent modes are excited, covers only three radial orders.
In this region, the distribution of frequencies is not random, the nonradial modes tend to cluster in three groups.
Since the parent frequencies tend to cluster in groups separated by
one radial order, the combinations, a$f_i\pm$b$f_j$, will also have to form clusters. Because even triple
combinations have been detected, the apparent semi-regular spacing over an enormous frequency range of 15 (!) radial orders is easily explained.

After eliminating the combination modes and harmonics, we are left with a pattern of three groups. Two of these groups cluster around the identified frequencies of the radial fundamental and first overtone, 6.90 and 8.96 cd$^{-1}$, respectively. Although only two radial modes have been identified in
the data, the stellar modeling (Lenz et al. 2008) allows us to predict the frequency of the second radial overtone at 11.21 cd$^{-1}$. This frequency is located near the third group. 

\begin{figure}
\centering
\includegraphics[bb=20 440 540 750,width=85mm,clip]{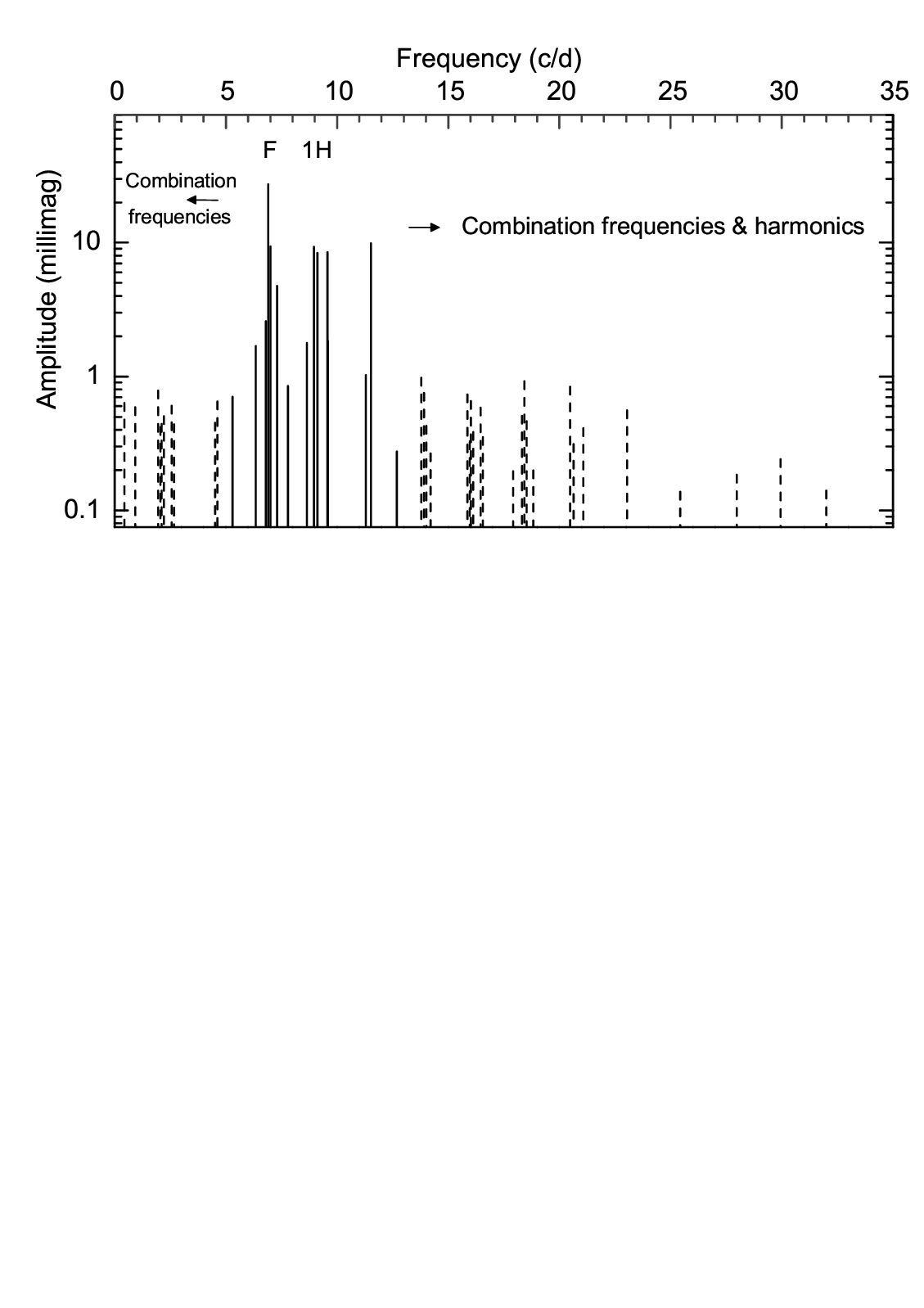}
\caption{Distribution of detected frequencies in 44 Tau. Because of the wide range in amplitudes, the amplitudes are plotted logarithmically. This diagram demonstrates that the detected frequencies show a semi-regular pattern. The main excitation range lies between 5.3 and 12.7 cd$^{-1}$. Frequencies outside this range were found to be either combination frequencies or harmonics and are marked by dashed lines.}
\end{figure}

\subsection{BL Cam}

The high-amplitude, extremely metal-deficient variable BL~Cam was investigated by Rodr\'iguez et al. (2007). The
authors identified 25 frequencies, of which 22 represent independent modes. 
The study is remarkable because of the difficulty of detecting such a large number of small-amplitude nonradial modes
in the presence of a dominant radial fundamental mode of high amplitude.

For statistics to examine regularities, the number of frequencies known for this star is fairly low. However, the authors note that
the frequencies of the nonradial modes cluster in groups near 25, 32, 46, 51--53, and 72--80 cd$^{-1}$. Since for this star successive radial orders are separated by about 7 cd$^{-1}$,
this suggests a possible similarity to the behavior of 44 Tau.

Following the evidence given by Rodr\'iguez et al. (2007), we have accepted the 25.576 cd$^{-1}$ frequency as the radial fundamental mode and based our
pulsation model on this identification. We have then computed the frequencies of the radial overtones. The results were similar to those presented
by Rodr\'iguez et al. (2007), e.g., the first overtone was predicted to be at 32.65 cd$^{-1}$. In fact, we confirm that in this star, the nonradial modes tend to
occur near (but not at) the predicted frequencies of the radial modes.

\subsection{FG Vir}

FG Vir is probably the best $\delta$~Scuti star to examine systematics in the pulsation frequencies. Omitting known combination frequencies and
harmonics leaves us with 68 frequencies. Contrary to the situation in 44 Tau, these independent frequencies cover a wide range from 5.7 to 44 cd$^{-1}$.
Visual inspection of the distribution (see Fig. 1) reveals that these frequencies are not distributed at random.
The regions with most frequencies are around 12, 23 and 34 cd$^{-1}$, but the data contain considerably more information than a spacing of 11 cd$^{-1}$. Note that 11 cd$^{-1}$ corresponds to approximately three radial orders.

In order to investigate the regularities of the frequency spacings  in a more systematic way we have applied the following simple scheme: we form frequency
differences between all the known frequencies and examine the histograms of the differences. This is  a sensitive test for patterns without statistical assumptions. We have carried out the analysis for all 80 frequencies as well as for only the independent frequencies. A specific pattern with peaks spaced $\sim 3.7$ cd$^{-1}$ apart shows up in both histograms. 
In practice, the harmonics and combination frequencies are easily recognized by their values and consequently can be omitted. Fig.~3 shows the pattern of the independent frequencies with the peaks denoted by arrows. We note from our models for FG Vir that the spacing of 3.7 cd$^{-1}$ corresponds to the average spacing between consecutive radial orders. Since the modes are almost all nonradial, we conclude that the most common separation of the photometrically visible nonradial
modes is, in fact, the same as that of the radial modes.

\begin{figure}
\centering
\includegraphics[bb=30 370 495 700,width=85mm,clip]{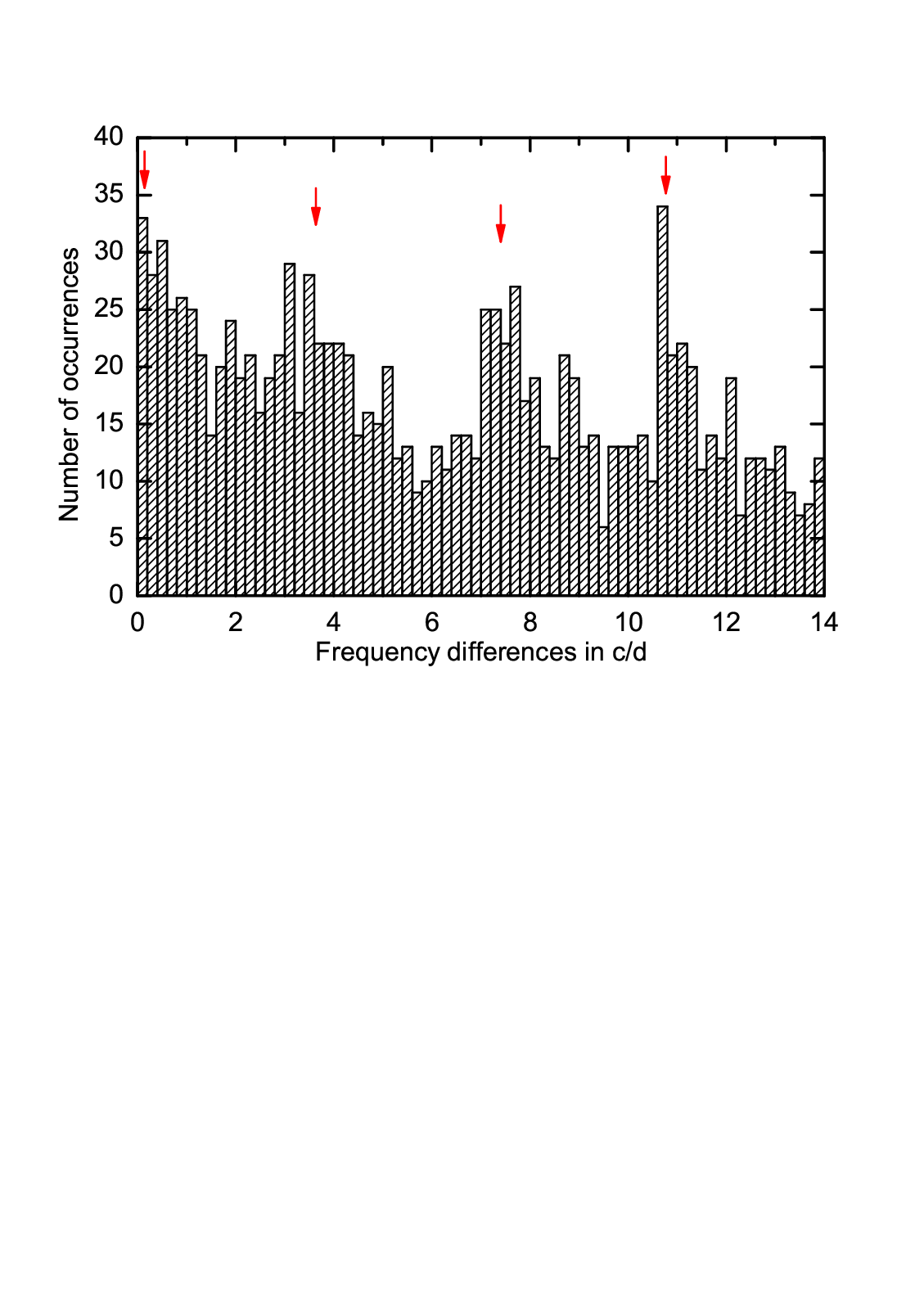}
\caption{Histogram of the frequency differences in FG Vir. Here every detected frequency was subtracted from all other frequencies. Known harmonics and combination
modes were omitted. The diagram shows that there
exists a preferred spacing (arrows) between the nonradial modes, which corresponds to the spacing between radial modes.}
\label{fig:obshisto}
\end{figure}

\subsection{Close frequency pairs}

For a number of $\delta$~Scuti stars, the observed amplitude and phase variability with time scales less than $\sim$200 d
is caused by beating between close frequencies, rather than true amplitude and period variability (Breger \& Bischof 2002, Breger \& Pamyatnykh 2006). This
was shown by the synchronous variation of amplitudes and phases, faithfully repeated in successive beat cycles. It was also found that, in general,
close frequencies are common in $\delta$~Scuti stars. Furthermore, the close frequencies show a preference to be close to those of radial modes. The number
of close frequencies is much higher than would be predicted by random coincidences and the astrophysical reason for such behaviour remains unclear.

\section{Frequencies of nonradial modes near those of radial modes}

The previous sections have shown that for the $\delta$~Scuti stars examined in detail, the frequencies of the nonradial modes are not distributed at random but show a preferred spacing corresponding to that of the radial modes. We wish to emphasize that this is only a preferred spacing and that other spacings do (and should) occur. 
Note that the evolutionary stage is different for these stars. While FG Vir and BL Cam are main-sequence stars, 44 Tau is already more evolved.

The question arises where in the frequency spectrum these concentrations of nonradial modes occur. Pulsation mode identifications for FG Vir might suggest that all visible modes cluster around specific values: the three close modes at 12.15, 12.72 and 12.79 cd$^{-1}$ have $\ell$ values of 0, 1 and 2, respectively (Breger et al. 1999, Daszy\'nska-Daszkiewicz et al. 2005, Zima et al. 2006).

We know that the vast majority of these observed modes cannot be radial modes, since they are much too numerous. In $\delta$~Scuti stars, mode identifications are available for only a small fraction of the known nonradial modes. We have examined how close the frequencies of these
modes are to the frequencies of the known or expected radial modes for the three stars examined above. The first step involved the computation of the radial
frequencies for these stars.

For our computations we used the same codes as described in Lenz et al. (2008). In most of the stars discussed in this paper a radial mode is observed. Consequently, after the determination of the radial order the mean density of the star is known. In the most cases rotation rates and metallicity have been determined from spectroscopy. This makes it possible to predict the position of other radial modes to a sufficient accuracy for studying the clustering of frequencies.
We have then computed the frequency difference of each detected mode to the nearest radial mode.

A difficulty of combining these data over large frequency ranges concerns the fact that we are not in the asymptotic range. This means that the frequency differences between successive radial orders are not constant. In an extreme case, for FG Vir, the frequency difference between the radial fundamental and the first overtone is
3.48~cd$^{-1}$, while for the 8th and 9th overtones the difference becomes 3.98~cd$^{-1}$. This effect degrades the patterns in the histograms, but cannot destroy them.

The histograms of the frequency differences are shown in Fig. 4. While in Fig. 4 the frequency distances are given in absolute values, in Fig. 5 the symmetry in the distribution of frequency differences for FG Vir is shown. The results are striking:
the frequencies of the photometrically detected nonradial modes are not distributed at random but tend to cluster around those of the radial modes. These nonradial modes are mostly $\ell$ = 1 and 2 modes.

\begin{figure*}
\centering
\includegraphics[bb=20 475 520 725,width=175mm,clip]{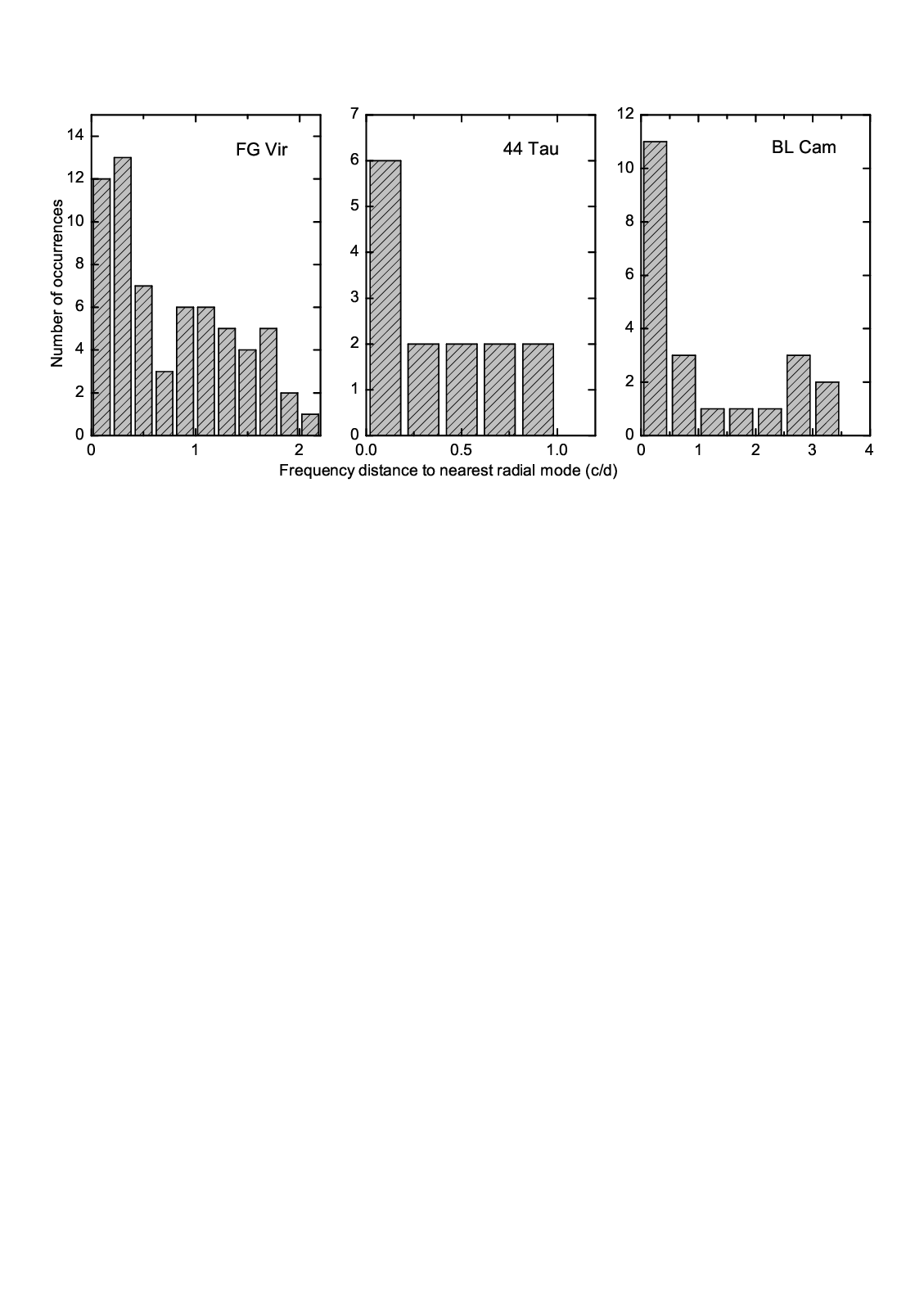}
\caption{Histogram of the frequency distances of individual modes to the frequency of the nearest radial mode. The frequencies of the radial modes were either
observed or computed from models. The range of frequency distances of each panel corresponds to approximately one radial order. This diagram shows that the observed nonradial modes are not distributed at random but tend to cluster around the radial modes.}
\label{fig:fig3}
\end{figure*}

\begin{figure}
\centering
\includegraphics[bb=35 520 520 780,width=85mm,clip]{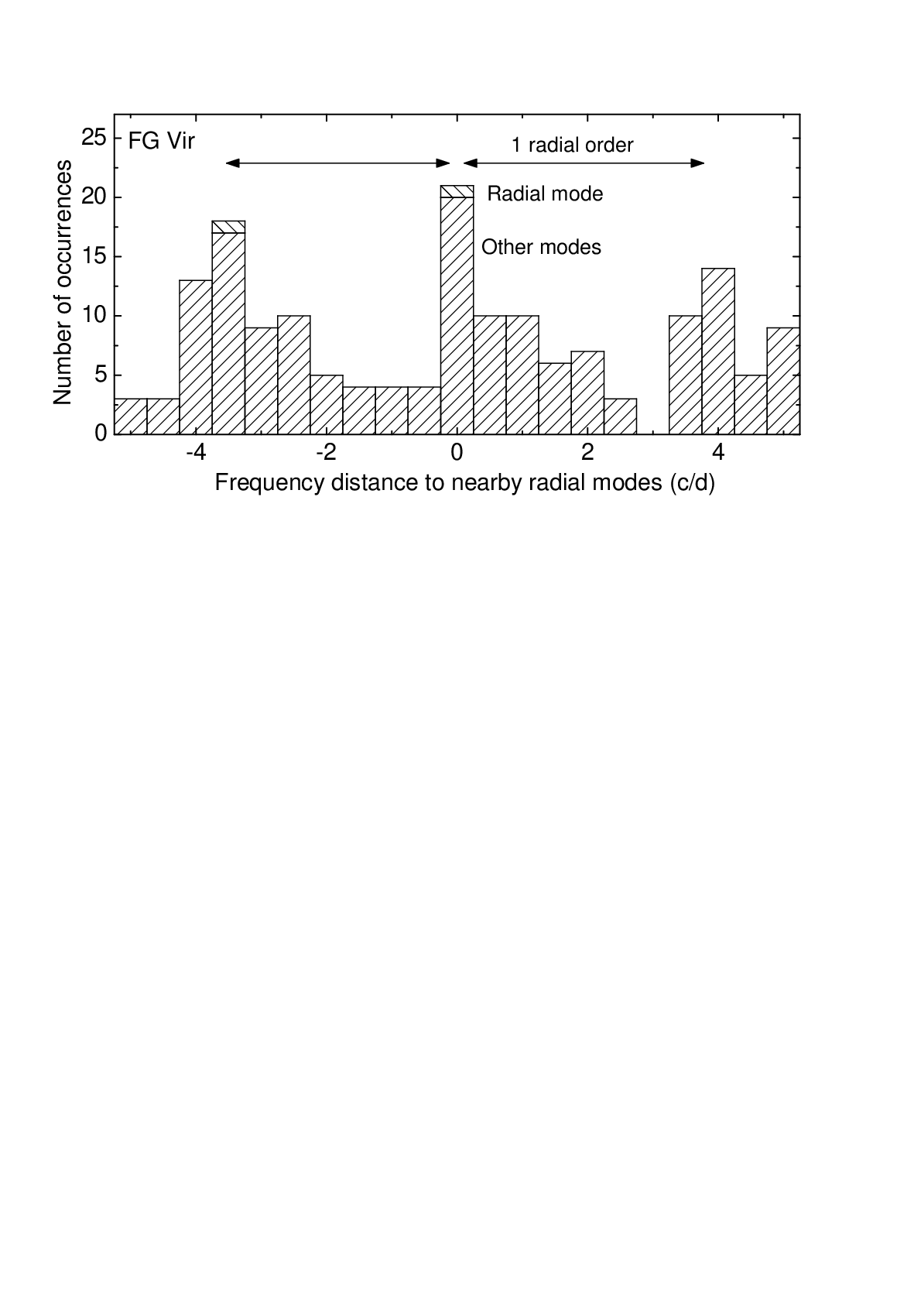}
\caption{Expanded histogram of the frequency distances of individual modes to the frequency of near radial modes in FG Vir. This diagram shows that the preferential clustering around the radial modes occurs in several radial orders.}
\label{fig:fig4}
\end{figure}

\section{Other stars}

The statistical analyses of the nonradial-mode frequencies requires the knowledge of a relatively large number of frequencies. These are available for only a few $\delta$~Scuti stars: three additional stars might be considered.

\subsection{$\epsilon$ Cep}

The main-sequence star $\epsilon$ Cep was studied with the WIRE satellite. 26 frequencies were reported (Bruntt et al. 2007). These authors note only weak clusters with spacings near 2.4, 1.2 cd$^{-1}$, 5.0 (or 5.8)  cd$^{-1}$ and compare these to the theoretical Large Spacing near 5 cd$^{-1}$. In the asymptotic limit of high-order acoustic modes, the Large Spacing defines the separation of frequencies of consecutive overtones of the same spherical harmonic degree, $\ell$. This separation is approximately equal to the spacing of radial modes. 

In the absence of knowledge of the frequencies of the radial modes, we have again calculated a histogram of
all the frequency differences, $f_i - f_j$, for the 22 independent frequencies. The largest, but not unique, feature is a spacing of broad peaks separated by 5 and 6 cd$^{-1}$.
This agrees with the spacing between successive radial modes, as estimated from the models of Bruntt et al. (2007).

\subsection{BI CMi}

BI CMi (Breger et al. 2002) shows 21 independent frequencies spread over a large number of orders. The frequency distribution does not show obvious patterns. Also, the stellar
parameters are uncertain so that we cannot predict the frequencies of the radial modes. This means that the radial proximity test (such as in Fig. 5)
cannot be applied.

\subsection{XX Pyx}

Handler et al. (2000) notes regularities in the spacing of the frequencies of the $\delta$~Scuti star XX Pyx, for which 22 frequencies were found. We also find the regularities in their data, although the three stars shown in the present paper are more convincing, due to their richer frequency spectra. 

\subsection{Other types of pulsators}

Recent results by Baran et al. (2008) showed that a similar frequency pattern as for $\delta$~Scuti stars is also seen in the hot subdwarf B star Balloon 090100001. They find that the observed p-modes form four distinct clusters. One frequency in the lowest-frequency cluster could be identified as a radial mode. Since the position of the other observed clusters matches the theoretically computed radial overtones, they find evidence that the groups of p-mode frequencies appear to lie in the vicinity of consecutive radial overtones.

\section{Mode trapping as an explanation for frequency clustering}

Dziembowski \& Kr\'{o}likowska (1990) examined mode trapping as a mechanism for mode selection in $\delta$~Scuti stars. They show that some modes of $\ell = 1$ are trapped in the envelope and, therefore, are less coupled to g modes in the deep interior. Such modes have a higher probability to be excited to observable amplitudes than other modes. Trapped modes are nonradial counterparts of the acoustic radial modes and, at low spherical degrees, their frequencies are close to those of radial modes. Indeed, in Lenz et al. (2008) the clustering of the observed frequencies in 44 Tau was shown to be in agreement with the theoretically predicted position of trapped modes.

In our linear computations we use the same normalization for all modes by adopting the relative amplitude of the radial displacement on the stellar surface to be equal 1. With such a normalization, modes trapped in the stellar envelope have a lower kinetic energy, $E_k$, than other modes. Furthermore, the fraction of kinetic energy from the acoustic cavity, $E_g$, is larger than for non-trapped modes, which results in minima in a $E_{g}/E_{k}$ diagram.

\begin{figure}
\centering
\includegraphics*[bb=15 10 260 225,width=80mm]{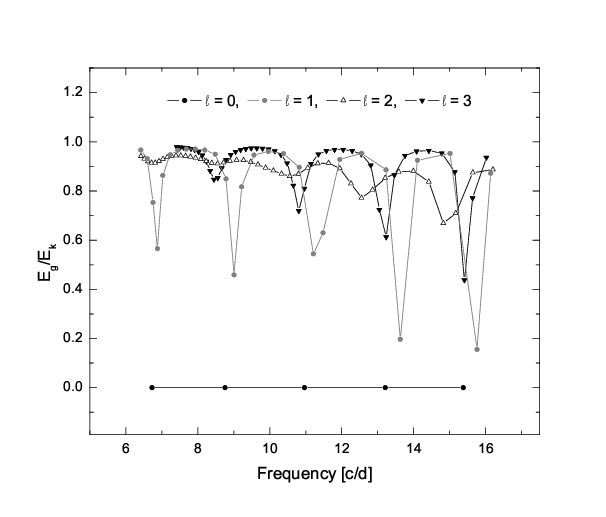}
\caption{Ratio of the kinetic energy from the gravitational cavity, $E_g$, to total kinetic energy, $E_k$, for unstable modes of $\ell \leq$ 3 for a post-main sequence $\delta$~Scuti model of 1.8 M$_{\odot}$. Modes trapped in the acoustic cavity are situated at local minima in this diagram.
}
\label{fig:ekin}
\end{figure}

In Fig.~\ref{fig:ekin} $E_{g}/E_{k}$ is given for unstable modes up to a spherical degree, $\ell$, of 3 for a post-main sequence $\delta$~Scuti model with a mass of 1.8 M$_{\odot}$. It can be seen that modes trapped in the envelope are situated close in frequency to the radial modes. 

Theory predicts that in $\delta$~Scuti stars mode trapping is marginal for $\ell$ = 2 modes (see also Fig.~\ref{fig:ekin}). However, our observations show that modes with $\ell$ = 2 occur preferentially around radial modes, e.g., in 44~Tau. It is difficult to explain why such a marginal effect of mode trapping can be sufficient for mode selection. Maybe another mode selection mechanism is effective for these modes.

A closer examination of modes with spherical harmonic degrees, $\ell$, higher than 3 reveals that with increasing $\ell$ only modes trapped in the envelope are unstable while other modes become stable. Photometrically, we mainly see modes with $\ell \leq 4$, although small-amplitude peaks are likely to arise from modes with higher spherical degrees as shown by Daszy\'nska-Daszkiewicz, Dziembowski \& Pamyatnykh (2006).

With increasing spherical degree, the trapped nonradial modes are located at higher frequencies than those of the radial modes. This effect can be also seen in observations, for example in the histograms of FG~Vir  (Fig. 5) and 44~Tau, which show a higher number of modes to the right of the central peak. It is not seen for BL Cam, but this may be due to a lower number of detected modes for this star.

An important effect that influences the position of nonradial modes is the avoided crossing phenomenon (Aizenman, Smeyers \& Weigert 1977). However, since avoided crossings of low-order modes only occur near the frequency values of radial modes, the pattern of the low-order frequencies is not distorted.

\section{Rotational Effects}

It is also necessary to study the impact of rotation on the clustering of frequencies around radial modes. We note:

(i) There exists an observational selection: the best-studied stars have less than average rotation. For the extreme case of 44 Tau with V$_{\rm rot}$ = 3 $\pm$ 2 km~s$^{-1}$ (Zima et al. 2006), the predicted rotational splitting is 0.02 cd$^{-1}$ or lower, much less than the separation of 2 cd$^{-1}$ for radial modes. In FG~Vir the observed rotational frequency separation is 0.53 cd$^{-1}$, while the spacing for each successive radial order is 4 cd$^{-1}$. Consequently, for this star modes with $m = \pm 2$ already degrade the histograms of frequency differences. However, our observations show that the regularities are still clearly visible in FG~Vir. 

(ii) Photometrically, so far we mostly observed axisymmetric modes (e.g., in FG Vir and 44 Tau). However, the other components may be found if the detection amplitude limit is lowered.

If these arguments are correct, we would predict that the groupings will become less obvious when the rotational splitting is of similar size as half the frequency difference between adjacent radial orders. 

To show the impact of rotation on the observational histograms presented in this paper, we computed pulsation modes up to $\ell = 2$ for a FG Vir model with a rotation rate of $V_{\rm rot}$~=~0 km~s$^{-1}$ and a model with $V_{\rm rot}$~=~136.7 km~s$^{-1}$. Only modes at minima in $E_g/E_k$ were considered. The rotational splitting for $\ell = 1$ and $\ell = 2$ modes was computed following the perturbation approach taking into account rotational effects up to second order (Dziembowski \& Goode 1992).

\begin{figure}
\centering
\includegraphics*[bb=10 15 310 230,width=88mm]{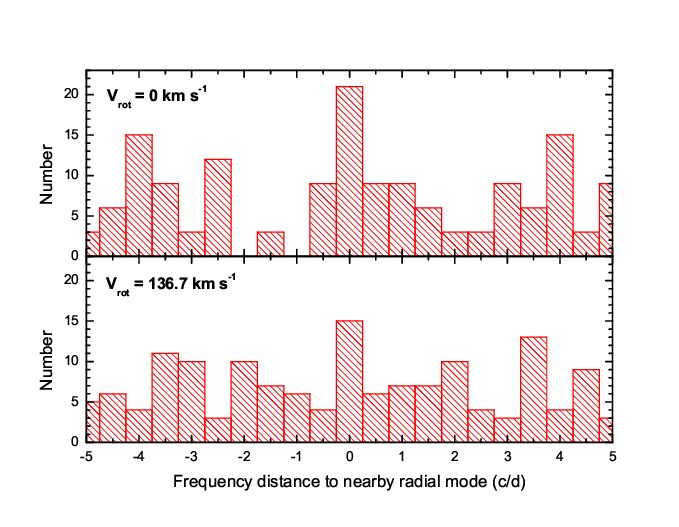}
\caption{Histogram of the frequency differences of predicted modes with $\ell$ = 0 to 2 to the nearby radial modes. Upper panel: rotationally split modes for $V_{\rm rot}$~=~0 km~s$^{-1}$. Lower panel: $V_{\rm rot}$~=~136.7 km~s$^{-1}$. Only unstable modes at minima in $E_g/E_k$ were considered (see Fig.~\ref{fig:ekin}). The structure is degraded by rotation, but still visible.
}
\label{fig:histo_vrot}
\end{figure}

The corresponding histograms are given in Fig.~\ref{fig:histo_vrot}. It can be seen that the clustering around radial modes becomes less pronounced for a higher rotation rate (lower panel). However, the structure is still present. 
The pattern in the histogram for fast rotation can be partly explained by different spacing of rotationally split modes for g~modes and p~modes. The amount of rotational splitting depends on  the value of the Ledoux Constant, C. C is approximately 0.5 for $\ell = 1$ g~modes and has a very small value for p~modes. Therefore, the rotational splitting for $\ell = 1$ g~modes is approximately half of that for the $\ell = 1$ p modes. Therefore, the structure seen in histograms for slow rotation is flattened when moving to fast rotation.

Another effect is the asymmetry of the rotational splitting that arises when including second-order terms. This effect is stronger for the p modes, which propagate in the envelope, and, therefore, are more strongly affected by the effects of rotational nonspherical distortion. 

Both effects lead to a smearing of the spacings of the rotationally split modes. We conclude that for rotation rates higher than approximately 100 km~s$^{-1}$, the regularities may become less pronounced, or even destroyed.

\section{Application of Regularities: The s-f diagram}

\begin{figure}
\centering
\includegraphics[bb=40 30 820 1150,width=88mm]{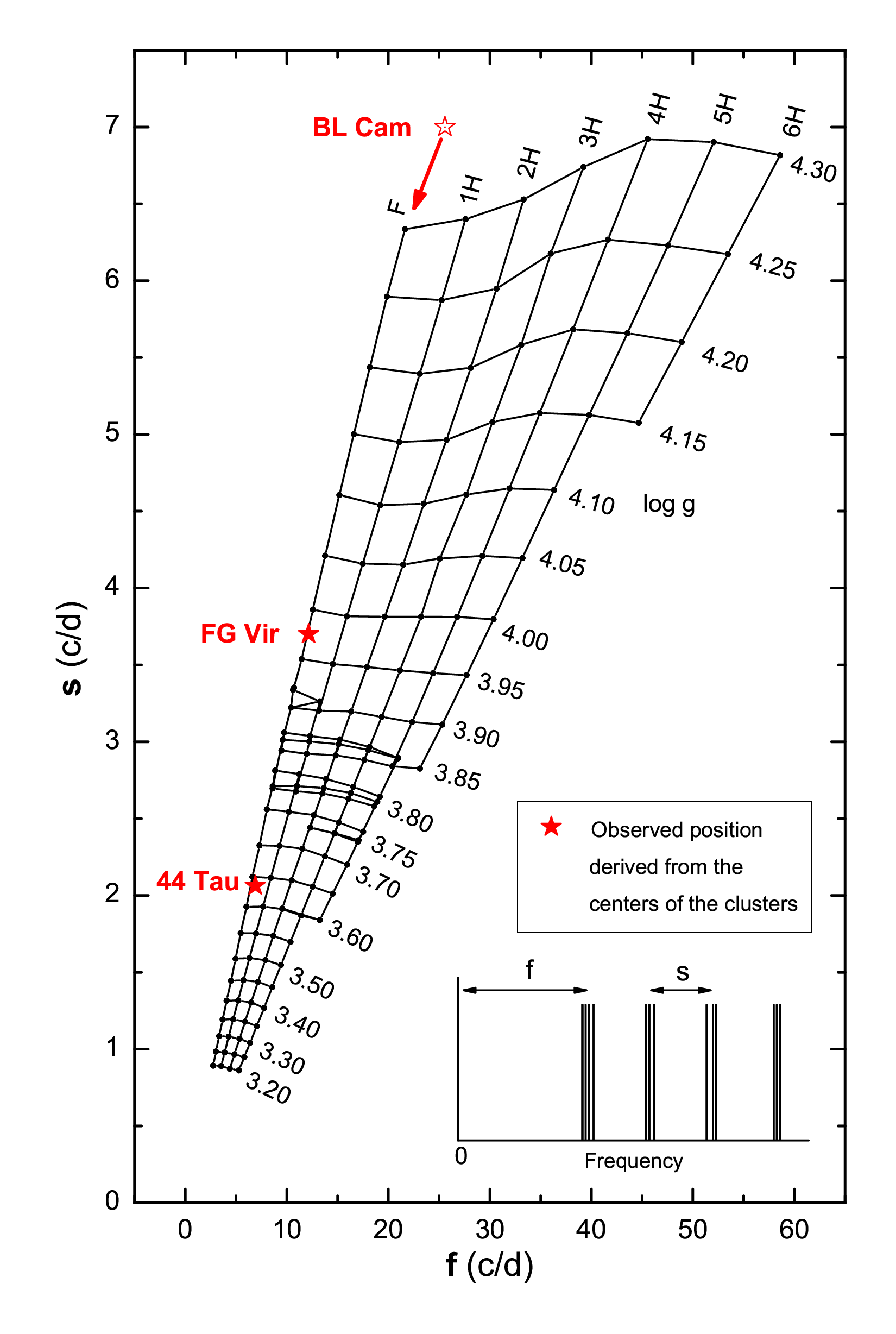}
\caption{Average separation of the radial modes, $s$, against the frequency of the lowest-frequency radial mode, $f$. We assume that each unstable radial mode represents the center of a cluster as illustrated on the bottom right. The grid makes it possible to determine $\log g$ and the order of the radial mode corresponding to the lowest frequency cluster. At $\log g$ = 3.75 to 3.90 a transition between main-sequence models and post-main sequence models takes place. The asterisks correspond to observed values for the cluster centers of 44 Tau, FG Vir and BL Cam.
}
\label{fig:grid}
\end{figure}

Following the observational results, we shall now assume that the centers of the observed frequency clusters correspond to the frequencies of the radial modes.
We will now show that by applying this assumption the presence of regularities in observed frequency spectra may be used to infer fundamental parameters of stars if these parameters are uncertain or unknown.

We computed a grid of models in the $\delta$~Scuti mass range (with $\Delta M$ = 0.025 M$_\odot$) using the Warsaw-New Jersey evolution code and Dziembowski's pulsation code. A short description of these codes is given in Lenz et al. (2008). 

Along each evolutionary track we computed the change of the radial frequencies. A detailed inspection of the pulsation models revealed an excellent possibility to determine the $\log g$ value of a star by means of two parameters: the average frequency separation between the radial modes, $s$, and the frequency of the lowest-frequency unstable radial mode, $f$.

\begin{figure}
\centering
\includegraphics*[bb=10 20 800 600,width=88mm]{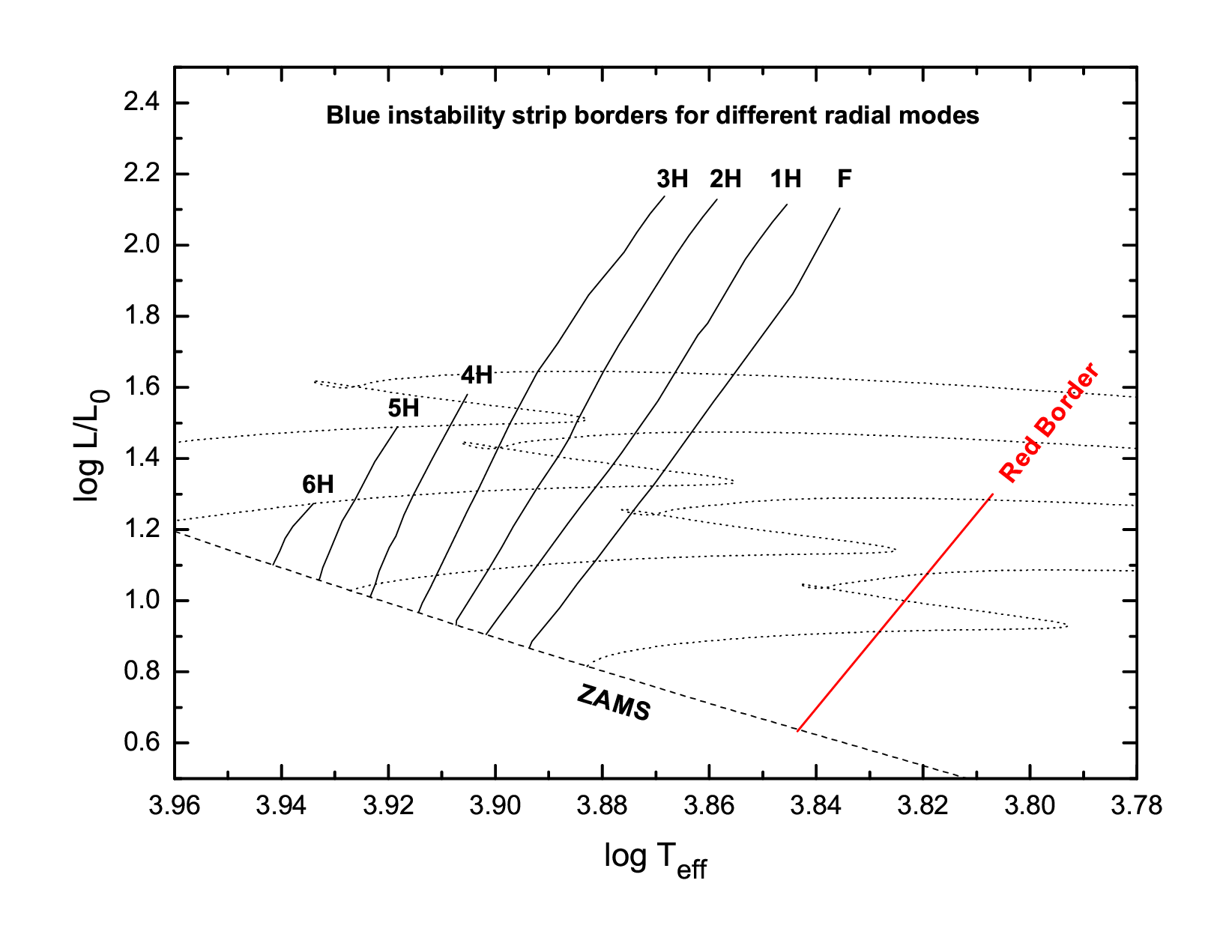}
\caption{Predicted blue borders of the instability strip. Evolutionary tracks for models with 1.6, 1.8, 2.0 and 2.2 M$_{\odot}$ are given by dotted lines. The red edge was taken from Dupret et al. (2004).
}
\label{fig:hrd_instab}
\end{figure}

Mode instability calculations show that the lowest-frequency cluster, $f$, corresponds to the position of the radial fundamental mode only for the cool $\delta$~Scuti stars. For the hotter stars, instability shifts to higher radial orders. The location of blue instability borders for modes up to the sixth radial overtone in a Hertzsprung-Russell Diagram (hereafter called HRD) is given in Fig.~\ref{fig:hrd_instab}. While the values of the $s$ and $f$ parameter depend on the mean density and, therefore, the evolutionary stage of a star, the $f$ value also includes a temperature dependence.

Fig.~\ref{fig:grid} shows the grid that was constructed for the $\log g$ determination. The grid points were derived from the positions of unstable radial modes. For each evolutionary track the model parameters and radial frequencies were interpolated between two time steps to accurately match the $\log g$ values shown in the grid. The $s$ value was then derived from the average spacing of all unstable radial modes, while $f$ corresponds to the frequency of the lowest unstable radial mode in the interpolated model. Finally, the $s$ and $f$ values from different stellar models with the same $\log g$ value were averaged to obtain the grid points. For this step, models cooler than the theoretical Red Edge given by Dupret et al. (2004) were rejected.

The observed positions derived from the centers of the clusters are shown for FG~Vir, 44~Tau and BL~Cam. The grid also provides information about the order of the radial mode that corresponds to the first cluster in the observed frequency spectrum. The s-f diagram  is in some respect similar to Petersen diagrams (period ratios of consecutive overtones versus the longer period of each pair). Petersen diagrams also allow to determine the order of observed radial modes (see, e.g., Olech et al. 2005, Fig. 6 there).

A transition between main-sequence models and post-main sequence models takes place at $\log g$ values between 3.90 and 3.75. During the main-sequence stage
the $\log g$ value decreases due to the slow expansion of the star. After the TAMS, the star contracts and its temperature and $\log g$ values increase again until the second turning point is reached; then the effective temperature and $\log g$ start to decrease again. For this reason, during the evolution of a star a $\log g$ value of 3.80, for example, can be reached three times. Due to the different stellar structure the radial frequency separation is slightly different. Therefore, for the $\log g$ value of 3.80, three horizontal grid lines (instead of one) are shown. The lowest of the three grid lines represents main-sequence models and the uppermost grid line post-main sequence models after the second turning point in the HRD.

\subsection{Uncertainties and limitations of the s-f diagram}

\begin{figure}
\centering
\includegraphics*[bb=00 00 210 310,width=88mm]{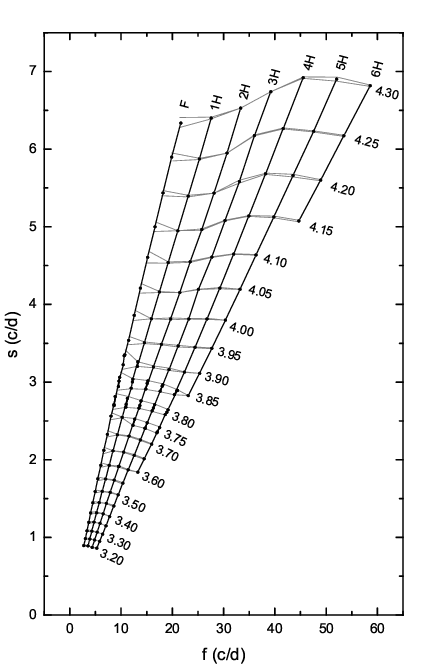}
\caption{Maximum and minimum value in $s$ (grey lines) for a given $\log g$ value in the domain of unstable modes (see Fig.~\ref{fig:hrd_instab}). The uncertainties in $\log g$ determination are highest when the radial fundamental mode is unstable. This is due to the larger range of model parameters such as effective temperature.
}
\label{fig:sfgriderr}
\end{figure}

\begin{figure}
\centering
\includegraphics*[bb=00 00 300 230,width=88mm]{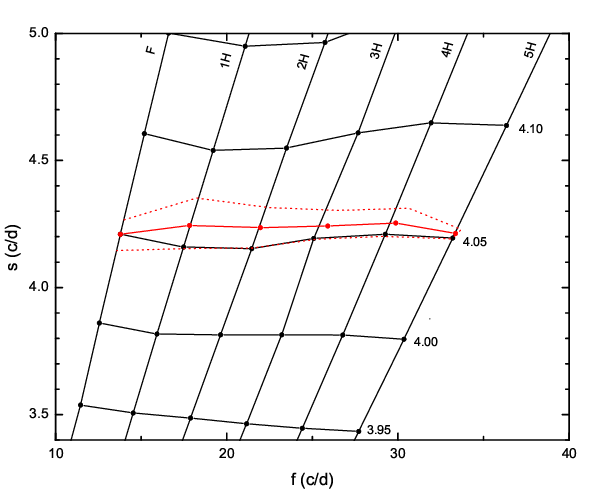}
\caption{If the lowest-frequency cluster and its corresponding radial mode are missed, the averaged grid points move to higher values of $s$. The new maximum and minimum values in $s$ for a $\log g$ value of 4.05 are marked by dotted lines. The uncertainties in the determination of $\log g$ at higher overtones are significantly higher as in Fig.~\ref{fig:sfgriderr}.
}
\label{fig:gridext1}
\end{figure}

We will now examine the uncertainties of the grid in detail. In the HRD, models with the same $\log g$ values are located almost parallel to the ZAMS. Models with different masses and effective temperatures were used to obtain average values of $s$ and $f$. As can be seen in the HRD in Fig.~\ref{fig:hrd_instab}, the models, for which the fundamental mode is predicted to be the lowest unstable radial mode, span the largest temperature and mass range. The first overtone is the predicted lowest unstable radial mode for models between the blue edge of the first radial overtone and the blue edge of the radial fundamental mode (only models located inside this region were used to compute the grid points marked as 1H). Therefore, the uncertainties in the case of the radial fundamental are larger than for the cases in which the lowest unstable radial mode is of higher order. In Fig.~\ref{fig:sfgriderr}, these uncertainties are given. The uncertainties of the grid in $\log g$ amount to $\leq$~0.05. The highest uncertainties exist for stars in the transition zone between the main sequence and post-main sequence models in which the fundamental radial mode is the lowest-frequency unstable mode.

The identification of frequency clusters in observational data may sometimes be difficult. The uncertainties given in Fig.~\ref{fig:sfgriderr} are based on the assumption that the cluster of the lowest-frequency unstable radial mode was correctly identified. What happens if the lowest-frequency cluster (corresponding to the lowest unstable radial mode) is missed, and we observe the cluster which corresponds to the next unstable radial overtone? We determined the uncertainties for the case when the lowest-frequency cluster is missed. The grid points slightly change and the uncertainties in $f$ and $s$ increase. This effect is shown in Fig.~\ref{fig:gridext1} where the corresponding uncertainties are marked with dotted lines. The deviation in $\log g$ may be as high as 0.03.

\begin{figure*}
\centering
\includegraphics*[bb=15 15 300 180,width=180mm]{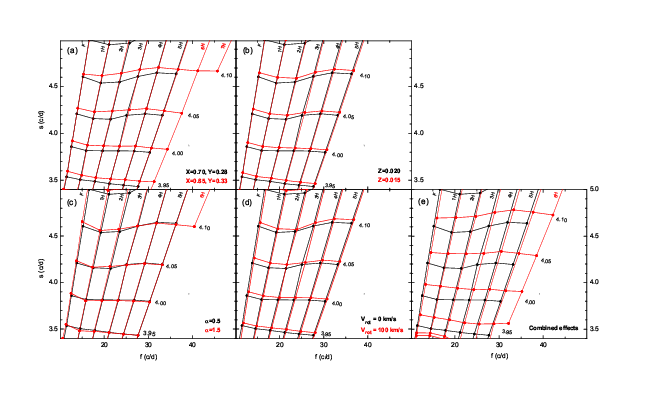}
\caption{The effects of different input parameters on the grid: (a) the effect of changing the helium abundance, (b) the effect of changing the metallicity from Z = 0.020 to Z = 0.015. Due to increased instability the grid is extended to higher radial orders. Panel (c) shows the effect of changing the efficiency of convection, (d) the effect of changing the equatorial rotational velocity from 0 to 100 km~s$^{-1}$. Rotational effects are considered up to second order. Panel (e) shows the effect of changing Y, Z and the equatorial rotational velocity by the amounts given in (a), (b) and (d) simultaneously.
}
\label{fig:gridtest}
\end{figure*}

The grid presented in this paper was obtained for nonrotating stellar models computed with the standard values for chemical composition (X = 0.70, Z = 0.02) using the GN93 element mixture and OPAL opacities. A mixing length parameter, $\alpha$, of 0.5 was used.
Since mode instability and the frequencies of the radial modes are affected by changes in helium abundance, metallicity, convective efficiency and rotation, we also computed corresponding models to test these effects. The results for different metallicity and helium abundance are shown in Fig.~\ref{fig:gridtest}(a) and~\ref{fig:gridtest}(b). The impact of a higher efficiency of convection on the grid is given in Fig.~\ref{fig:gridtest}(c). The effect of rotation on the grid is also small as long as rotation rates below 100~km~s$^{-1}$ are considered (see Fig.~\ref{fig:gridtest}(d)). Fig.~\ref{fig:gridtest}(e) shows the effect of changing Y, Z and the equatorial rotational velocity by the amounts given in (a), (b) and (d) simultaneously. Even here the deviations in $\log g$ are smaller than 0.03.  

In Fig.~\ref{fig:grid} we also show the observed positions of three stars in the s-f diagram. FG Vir and 44 Tau have normal chemical composition, whereas BL~Cam is extremely metal-deficient (Z = 0.0001). We computed pulsation models for BL~Cam to determine the shift in $\log g$. The amount of the shift is shown in Fig.~\ref{fig:grid} by an arrow, which may also be used to correct the $\log g$ value of other stars with very low metallicity. 

It is also vital to discuss the uncertainties of the observations and the assumptions that the radial mode is located at the center of a cluster and that the radial modes are equidistantly spaced. We have used the data for FG Vir and find that the uncertainties in $s$ are less than 10\%. The parameter $f$, on the other hand, necessarily leads
to a correct point on the grid, since only discrete position are possible (unless a complete order is missed, which is explored in Fig.~11). The observational uncertainties lead to a combined uncertainty of $\log g$ of 0.05.

In the preceeding sections we have shown that in the low-order frequency region the observed nonradial modes are situated close to the radial modes. In the asymptotic regime of high-order oscillations there is another common regular pattern that is also observed in the Sun. The $\ell = 1$ modes are located midway between $\ell = 0,2$ modes. The spacing of the frequency clusters changes from a full radial order at low frequencies to half-radial in the asymptotic frequency region.
Consider a star with an observed frequency separation of 2~cd$^{-1}$ and a first frequency cluster at 10 cd$^{-1}$. If we misinterpret this frequency difference to correspond to the half radial separation between $\ell$ = 0 and $\ell$ = 1 modes (as in the asymptotic case), the predicted radial separation would be 4 cd$^{-1}$. This value is located outside the grids shown in the s-f diagram, and no unstable modes are expected. Consequently, the incorrect value of 4 cd$^{-1}$ is ruled out so that 2 cd$^{-1}$ has to be the separation of radial frequencies. Any ambiguities may be ruled out this way and an incorrect $\log g$ determination is improbable.

Taking into account all the uncertainties given in this section, the determination of $\log g$ from the s-f diagram may provide the same or even better accuracy as from photometric data for stars with moderate rotational velocities and close to normal chemical abundances.

\section{Conclusions}

In the observed pulsation spectra of well-studied $\delta$~Scuti stars, such as 44~Tau, FG~Vir and BL Cam, a regular distribution of frequencies can be found. The detected frequencies tend to cluster in groups around radial modes. The comparison of the observations with theoretical pulsation models reveals that the cluster pattern may partly be explained by trapped modes in the stellar envelope. 

Following the assumption that the radial modes correspond to the center of the cluster, we construct the s-f diagram. It relates the two parameters $f$, the frequency of the lowest-frequency radial mode, and $s$, the mean spacing between the radial modes. Only linearly unstable modes are considered. The s-f diagram allows to infer the stellar $\log g$ value and to determine the order of the radial mode responsible for the lowest-frequency cluster for stars in the $\delta$~Scuti mass range. The diagram can be applied if clear patterns are seen in the frequency spectrum.
We examined the uncertainties of such a determination in detail and found them to be sufficiently small for a reliable $\log g$ determination of slowly or moderately rotating stars without chemical peculiarities.

\section*{Acknowledgements}

This investigation has been supported
by the Austrian Fonds zur F\"{o}rderung der wissenschaftlichen Forschung.


\begin{thebibliography}{}

\bibitem{} Aizenman M., Smeyers P., Weigert A., 1977, A\&A, 58, 41
\bibitem{} Baran A., Oreiro R., Pigulski A. et al., 2008, arXiv e-print (arXiv:0810.4010)
\bibitem{} Breger M., Bischof K. M., 2002, A\&A, 385, 537
\bibitem{} Breger M., Lenz P., 2008, A\&A, 488, 643
\bibitem{} Breger M., Pamyatnykh A. A., 2006, MNRAS, 368, 571
\bibitem{} Breger M., Pamyatnykh A. A., Pikall H., Garrido, R., 1999, A\&A, 341, 151
\bibitem{} Breger M., Garrido R., Handler G. et al., 2002, MNRAS, 329, 531
\bibitem{} Breger M., Lenz P., Antoci V. et al., 2005, A\&A, 435, 955
\bibitem{} Bruntt H., Su\'{a}rez J. C.; Bedding T. R. et al., 2007, A\&A, 461, 619
\bibitem{} Daszy\'nska-Daszkiewicz J., Dziembowski W.~A., Pamyatnykh A.~A., Breger M., Zima W., Houdek G., 2005, A\&A, 438, 653
\bibitem{} Daszy\'nska-Daszkiewicz J., Dziembowski W.~A., Pamyatnykh A.~A., 2006, MmSAI, 77, 113
\bibitem{} Dupret M.~A., Grigahc\'ene A., Garrido R., Gabriel M., Scuflaire R., 2004, A\&A, 414, 17
\bibitem{} Dziembowski W.~A., Goode P.~R., 1992, ApJ, 394, 670
\bibitem{} Dziembowski W.~A., Kr\'{o}likowska M., 1990, AcA, 40, 19
\bibitem{} Handler G., Arentoft T., Shobbrook R. R. et al., 2000, MNRAS, 318, 511
\bibitem{} Lenz P., Pamyatnykh A. A., Breger M., Antoci V., 2008, A\&A, 478, 855
\bibitem{} Olech A., Dziembowski W. A., Pamyatnykh A. A., Kaluzny J., Pych W., Schwarzenberg-Czerny A., Thompson I. B., 2005, MNRAS, 363, 40
\bibitem{} Rodr\'{i}guez E., Fauvaud S., Farrell J. A. et al., 2007, A\&A, 471, 255
\bibitem{} Stellingwerf R. F., 1979, ApJ, 227, 935
\bibitem{} Stellingwerf R. F., 1980, in ''Nonradial and nonlinear stellar pulsation'', Proc. Workshop (Tucson, 12-16 March 1979), Lecture Notes in Physics, Vol. 125, p. 50
\bibitem{} Zima W., Wright, D., Bentley, J. et al., 2006, A\&A 455, 235
\end{thebibliography}
\end{document}